# Artificial Intelligence for High-Throughput Discovery of Topological Insulators: the Example of Alloyed Tetradymites

Guohua Cao[1,2,3], Runhai Ouyang[2], Luca M. Ghiringhelli[2], Matthias Scheffler[2], Huijun Liu[1,*], Christian Carbogno[2,*], Zhenyu Zhang[3,*]

[1]*Key Laboratory of Artificial Micro- and Nano-Structures of Ministry of Education and School of Physics and Technology, Wuhan University, Wuhan 430072, China*

[2]*Fritz-Haber-Institut der Max-Planck-Gesellschaft, Faradayweg 4-6, 14195 Berlin-Dahlem, Germany*

[3]*International Center for Quantum Design of Functional Materials (ICQD), Hefei National Laboratory for Physical Sciences at the Microscale, and Synergetic Innovation Center of Quantum Information and Quantum Physics, University of Science and Technology of China, Hefei, Anhui 230026, China*

**ABSTRACT** Significant advances have been made in predicting new topological materials using high-throughput empirical descriptors or symmetry-based indicators. To date, these approaches have been applied to materials in existing databases, and are severely limited to systems with well-defined symmetries, leaving a much larger materials space unexplored. Using tetradymites as a prototypical class of examples, we uncover a novel two-dimensional descriptor by applying an artificial intelligence (AI) based approach for fast and reliable identification of the topological characters of a drastically expanded range of materials, without prior determination of their specific symmetries and detailed band structures. By leveraging this descriptor that contains only the atomic number and electronegativity of the constituent species, we have readily scanned a huge number of alloys in the tetradymite family. Strikingly, nearly half of which are identified to be topological insulators, revealing a much larger territory of the topological materials world. The present work also attests the increasingly important role of such AI-based approaches in modern materials discovery.

# I. INTRODUCTION

Topological insulators (TIs) constitute a new class of quantum materials with insulating bulk but metallic boundary states. Protected by time-reversal symmetry [1-3], those boundary states possess a spin-momentum locked Dirac structure in which the backscattering channels are suppressed, favoring dissipationless electronic conduction. This salient property renders TIs immense application potential in many areas, including spintronics [4,5], catalysis [6,7], and thermoelectricity [8], propelling the field to actively search for new TIs that may meet such high technological expectations.

Theoretically, the TIs are characterized by a non-vanishing topological invariant $Z_2$, which can be determined by parity considerations or by integration of the Berry curvature in momentum space through detailed electronic structure calculations [9-11]. Earlier search efforts for potential TIs were performed on a case-by-case basis, as exemplified by the successful predictions and experimental confirmations of the well-known strong TIs $Bi_2Te_3$, $Sb_2Te_3$, and $Bi_2Se_3$ [12-15]. More recently, the field has witnessed rapid advances in high-throughput screening over many candidate topological materials using empirical descriptors or symmetry-based indicators [16-22]. Out of the tens of thousands of materials with well-defined symmetries in the Inorganic Crystal Structure Database (ICSD), hundreds to thousands have been identified to be topologically nontrivial.

Beyond the systems listed in the existing databases, there exists a much larger materials space, e.g., compounds that can be created by alloying or fractionally varying the stoichiometric compositions. While these immensely vast material classes were out of the scope of the symmetry-based approaches, they are the focus of the present study. Using tetradymites as a prototypical class of examples, we establish an artificial intelligence (AI)-based descriptor for the prediction of TIs without prior determination of their specific symmetries and detailed band structures, covering a previously uncharted and much larger territory in the materials space. To this goal, we first investigate a moderate number (hundreds) of layered tetradymites by using high-level electronic structure calculations that account for van der Waals (vdW) interactions [23-25] and many-body effects [26,27] at a perturbative level. We then employ the SISSO (Sure Independence Screening and Sparsifying Operator) approach [28,29] based on the compressed-sensing technique to establish a simple and physically intuitive descriptor for the identification of the topological character. By leveraging this descriptor that contains only the atomic number and electronegativity of the constituent species, we have readily scanned a huge number of alloys in the tetradymite family. Strikingly, nearly half of which are identified to be topological insulators. The present work also attests the increasingly important role of such AI-based approaches in modern materials discovery.

## II. RESULTS AND DISCUSSION

In contrast to other AI-based approaches applied to the prediction of TIs [30−35], SISSO is built to determine descriptors that are elementary functions of key physical inputs, thus enabling human inspection into the underlying mechanisms. To construct a reliable training set, we have computed the topological characters of 243 (or $3^5$) tetradymites by combining group-VA elements (As, Sb, and Bi) with group-VIA elements (S, Se, and Te) in a five-atom unit cell. These systems can be viewed as stackings of quintuple layers (QLs) along the *c*-direction, where vdW interactions bind neighboring QLs to each other. As an example, Fig. 1 shows the crystal structure of the tetradymite AsSbSeTeS, where the atoms As, Sb, Se, Te, and S occupy the sites *A*, *B*, *L*, *M*, and *N*, respectively. Here we employ high-level first-principles methods including appropriate vdW-functionals and quasiparticle corrections as detailed in Section 1 of the Supplemental Material [36], which are essential to obtaining reliable values for the topological invariant $Z_2$.

Based on the high-level electronic structure data, we utilize SISSO [28,29] to single out a simple and physically intuitive descriptor from a huge number of potential candidate forms. In practice, more than ten billion candidate descriptors are first constructed iteratively by combining elemental physical properties of the constituent atoms. More details can be found in Section 1 of the Supplemental Material [36]. Secondly, the optimized descriptor is obtained via the SISSO approach and is used to determine a low-dimensional representation of the materials space, in which the TIs and the normal insulators (NIs) belong to well-defined, non-overlapping domains. Since the transparent functional form of the identified descriptor reflects the mechanism underlying the topological character, it is physically meaningful and suitable for extrapolation, as also demonstrated below.

By performing accurate first-principles calculations for the set of 243 teradymites, we identify 177 systems that are mechanically stable (namely, without pronounced negative phonons), while 66 are unstable. We also find that the topological characters of 230 systems do not depend on the special choice of the vdW functional, while the other 13 systems give different $Z_2$ values for the optB86b-vdW and optB88-vdW functionals mentioned in Section 1 of the Supplemental Material [36]. Furthermore, among the 177 stable systems, there are 10 weak TIs (namely, those with small band gaps, and their topological surface states only exist on certain surface orientations), and 4 semimetals. Excluding all these somewhat ambiguous systems, we finally obtain a set of 152 stable tetradymites with clear TI/NI classifications as the training set. Among them, 67 systems are TIs (including 4 binary, 27 ternary, 30 quaternary, and 6 quinary tetradymites), while the remaining 85 systems are NIs (containing 3 binary, 24 ternary, 46 quaternary, and 12 quinary tetradymites). The stability and topological properties of all the 243 tetradymites are summarized in Section 2 of the Supplemental Material [36]. It should be mentioned that we are dealing with the topological properties of intrinsic tetradymites. The effect of defects and doping, which may destroy the salient properties of TIs [14, 37, 38], is however beyond the focus of our current work.

It should be noted that the training data set naturally includes the TIs $Bi_2Te_3$, $Bi_2Se_3$, $Sb_2Te_3$, $Sb_2Te_2S$, $Sb_2Te_2Se$, $Bi_2Te_2Se$, $Bi_2Se_2Te$ and $BiSbSeTe_2$ that have already been previously identified [12-15,39-41]. Furthermore, the set also includes 59 new TIs, and some of those feature relatively large band gaps, such as SbBiSeSeSe (0.22 eV), SbBiTeTeS (0.27 eV), and SbBiTeTeSe (0.23 eV), and are thus promising candidates for potential applications. On the other hand, we want to emphasize that for a certain tetradymite, the topological nature may be sensitive to the atomic configuration within the QL. Taking $Sb_2Te_2S$ as an example, our additional first-principles calculations indicate that the structure with constituent elements arranged as SbSbTeTeS is identified to be TI, while SbSbSTeTe with the same stoichiometry is a NI. Even at this level, these new findings are significant.

The SISSO training has been performed for the described data set of 152 tetradymites by using different combinations of key physical inputs: (1) the atomic number $Z$, the electronegativity $\chi$, and the spin-orbit coupling (SOC) strength $\lambda$ [42]; (2) $\chi$ and $\lambda$; and (3) only $\lambda$. More details can be found in Section 3 of the Supplemental Material [36]. The optimized two-dimensional (2D) descriptor is identified as follows:

$$D_1 = (Z_A + Z_B) \cdot (Z_L + Z_M) - |Z_A Z_M - Z_B Z_L|, \tag{1}$$

$$D_2 = \left| \frac{(\chi_L + \chi_M) \cdot Z_N}{\chi_A} - (Z_L + Z_M) - |Z_L - Z_M| \right|. \tag{2}$$

Here, the subscripts $A$, $B$, $L$, $M$, and $N$ refer to the sites occupied by the atoms (see Fig. 1). We note that, even though $\lambda$ has been widely recognized to induce band inversion and topological phase transition, it was not explicitly selected by the optimized descriptor. Using $D_1$ and $D_2$ as the $x$ and $y$ coordinates, we can plot a "topological phase diagram" of the 152 tetradymites as shown in Fig. 2, where the cyan and green regions are the convex envelopes of the 85 NIs and 67 TIs, respectively. The support vector machine (SVM) technique [43] is further used to obtain the blue dividing line $D_2 = -238.23 + 0.039 D_1$ between the TI and NI domains. We see from Fig. 2 that the resulting 2D descriptor gives a perfect classification of the training data, i.e., there is no overlap between the NI and TI domains. The robustness of the descriptor has been checked via the leave-one-out cross-validation approach, and the all-data descriptor is identified in 86% of all the 152 iterations (each descriptor is identified out of ~$10^{23}$ possible candidates). Furthermore, it is remarkable that even the 66 mechanically unstable systems (open symbols) are perfectly classified by the phase diagram, despite the fact that the descriptor is obtained by using only the 152 stable ones (filled symbols) as the training set. Even more, exactly the same 2D descriptor ($D_1, D_2$) is identified if the full data of 216 stable and unstable tetradymites are adopted. These striking observations further substantiate that the obtained 2D descriptor is able to capture the underlying physical mechanism determining the topological characters in this materials class.

We can rewrite Eq. (1) as

$$D_1 = \begin{cases} Z_A Z_L + 2 Z_B Z_L + Z_B Z_M, & \text{if } Z_A Z_M - Z_B Z_L > 0 \\ Z_A Z_L + 2 Z_A Z_M + Z_B Z_M, & \text{if } Z_A Z_M - Z_B Z_L < 0 \end{cases}, \quad (3)$$

showing that larger values of $Z_A$, $Z_B$, $Z_L$ and $Z_M$ lead to higher $D_1$, making the system to be in the TI region. This is consistent with the common understanding that heavier atoms usually have larger SOC, which is a key factor in inducing topological band inversion [12]. For instance, the well-known TI Bi$_2$Te$_3$ exhibits the largest $D_1$ value among all the 152 tetradymites and thus represents a vertex of the convex TI envelope, since it features the largest atomic numbers of the cations ($Z_A = Z_B = 83$) and anions ($Z_L = Z_M = Z_N = 52$). It should also be noted that $Z_N$ does not appear in Eq. (3), because the anions at the $N$ sites make minimal contributions to the highest valence band and lowest conduction band of the systems. More details can be found in Section 4 of the Supplemental Material [36].

Here we stress that, though $D_1$ is the decisive descriptor for a large majority of these tetradymites, it alone is not sufficient to predict the topological character in the region with $6806 < D_1 < 7854$ (enclosed by two vertical lines in Fig. 2). In this case, $D_2$ of the 2D descriptor becomes crucial: among the 29 tetradymites located within this particular region, 14 compounds with smaller $D_2$ values are TIs, while the remaining 15 compounds with larger $D_2$ values are NIs. To better reveal the delicate physical mechanism, we rewrite Eq. (2) as

$$D_2 = \begin{cases} \left|\dfrac{(\chi_L + \chi_M)\cdot Z_N - 2\chi_A Z_L}{\chi_A}\right|, & \text{if } Z_L - Z_M > 0 \\ \left|\dfrac{(\chi_L + \chi_M)\cdot Z_N - 2\chi_A Z_M}{\chi_A}\right|, & \text{if } Z_L - Z_M < 0 \end{cases}. \quad (4)$$

Conceptually, $D_2$ is the relative electronegativity difference between the anions (at the $L$ and $M$ sites) and cations (at the $A$ sites). As the electronegativity difference is approximately positively correlated with the band gap of an inorganic compound [44,45], a smaller $D_2$ value (namely, a smaller electronegativity difference) leads to a smaller band gap, making it easier to generate a band inversion, and the system is more likely a TI. Indeed, our first-principles calculations for the 29 tetradymites (with $6806 < D_1 < 7854$) indicate that the TIs with smaller $D_2$ tend to exhibit smaller band gaps (before SOC) than the respective NIs at similar $D_1$. Details can be found in Section 4 of the Supplemental Material [36].

Overall, although the proposed 2D descriptor only depends on the atomic number $Z$ and the electronegativity $\chi$ of the constituent atoms, it correctly captures the essential physical factors of TIs, namely, the delicate competition between the SOC strength and band gap. Compared with previous empirical models [16,18], the present 2D descriptor of ($D_1, D_2$) identified by SISSO features a much richer functional form, enabling quantitative and reliable predictions of TIs far beyond the training data, as further demonstrated below.

The 243 tetradymites discussed so far all feature integer stoichiometry. In fact,

alloyed tetradymites with fractional stoichiometry such as $Bi_2Te_{3-x}S_x$ and $Bi_{1.4}Sb_{0.6}Te_{1.8}S_{1.2}$ [46] are promising TIs [47] for potential broader technological applications, since such variations allow to further tune the electronic and topological properties. Unfortunately, reliable first-principles data is hard if not impossible to obtain for such systems, since prohibitively large supercells are needed to represent such fractional stoichiometries given by $As_xSb_yBi_{2-x-y}S_aSe_bTe_{3-a-b}$ ( $0 \leq x, y \leq 2$ and $0 \leq a, b \leq 3$). Generalizing the 2D descriptor discovered above, however, allows to predict the topological characters for the complete class of tetradymites, including alloyed compounds. For this purpose, we define site-specific atomic numbers as:

$$\begin{aligned}
Z_A &= x_1 Z_{As} + y_1 Z_{Sb} + (1 - x_1 - y_1) Z_{Bi} \\
Z_B &= x_2 Z_{As} + y_2 Z_{Sb} + (1 - x_2 - y_2) Z_{Bi} \\
Z_L &= a_1 Z_S + b_1 Z_{Se} + (1 - a_1 - b_1) Z_{Te} \\
Z_M &= a_2 Z_S + b_2 Z_{Se} + (1 - a_2 - b_2) Z_{Te} \\
Z_N &= a_3 Z_S + b_3 Z_{Se} + (1 - a_3 - b_3) Z_{Te}
\end{aligned} \quad (5)$$

and site-specific electronegativities as:

$$\begin{aligned}
\chi_A &= x_1 \chi_{As} + y_1 \chi_{Sb} + (1 - x_1 - y_1) \chi_{Bi} \\
\chi_B &= x_2 \chi_{As} + y_2 \chi_{Sb} + (1 - x_2 - y_2) \chi_{Bi} \\
\chi_L &= a_1 \chi_S + b_1 \chi_{Se} + (1 - a_1 - b_1) \chi_{Te} \\
\chi_M &= a_2 \chi_S + b_2 \chi_{Se} + (1 - a_2 - b_2) \chi_{Te} \\
\chi_N &= a_3 \chi_S + b_3 \chi_{Se} + (1 - a_3 - b_3) \chi_{Te}
\end{aligned} \quad (6)$$

In close analogy to the virtual crystal approximation [48] used for first-principles modelling of random alloys, these site-specific properties are weighted averages, i.e., the coefficients $x_1, x_2, a_1, a_2, a_3, y_1, y_2, b_1, b_2, b_3$ (each taking the value between 0 and 1) denote the fractional occupancy of each site. Collectively, these coefficients define the stoichiometry of an alloyed system via $x = x_1 + x_2$, $y = y_1 + y_2$, $a = a_1 + a_2 + a_3$, and $b = b_1 + b_2 + b_3$.

Utilizing the weighted elemental properties to evaluate the descriptor in Eq. (1) and (2) immediately allows to map out the topological phase diagram for the tetradymites with any fractional stoichiometry. This is illustrated in Fig. 3 using 4,084,101 alloyed tetradymites as yet still only a subset of the example systems, where the coefficients $x_1, x_2, a_1, a_2, a_3, y_1, y_2, b_1, b_2, b_3$ are all multiples of 0.2. Even in this relatively small subset of alloyed tetradymites, we already obtain 1,965,047 systems located on the right side of the dividing line, i.e., they are predicted to be TIs.

Noteworthy enough, the obtained predictions are consistent with the few available experimental data, i.e., the topologically non-trivial alloyed tetradymites $Bi_2Te_{1.6}S_{1.4}$, $Bi_{0.5}Sb_{1.5}Te_3$, $Bi_{1.5}Sb_{0.5}Te_3$, $Bi_{1.5}Sb_{0.5}Se_{1.3}Te_{1.7}$, and $Bi_{1.1}Sb_{0.9}STe_2$ [40,47,49]. Taking $Bi_2Te_{1.6}S_{1.4}$ as an example, the atomic arrangement within the unit cell is Bi, Bi, $Te_{0.8}S_{0.2}$, $Te_{0.8}S_{0.2}$, S, occupying the sites $A$, $B$, $L$, $M$, $N$, respectively. The calculated ($D_1$, $D_2$) values are (14873.6, 54.6), which appears in the TI domain. This observation is not surprising, as the Bi- and Te-rich random alloys marked by the solid asterisks in Fig. 3 feature large atomic masses and thus large values of $D_1$. To validate the full topological phase diagram, we have further performed explicit first-principles calculations for eight additional tetradymites with fractional stoichiometry that are still reasonably tractable. Marked as the solid circles (NIs) and squares (TIs), these representative examples cover all the different areas shown in Fig. 3, including the regions close to the NI/TI boundary. Accordingly, four tetradymites sample the TI area ($As_{0.33}Sb_{0.67}BiSe_2Te$, $Sb_2S_{0.33}Se_{0.67}Te_2$, $Sb_{1.67}Bi_{0.33}S_{0.67}Te_{1.33}Se$ and $BiSbS_{0.67}SeTe_{1.33}$), while the other four are located within the NI area ($SbBiS_2Se_{0.67}Te_{0.33}$, $SbBiS_{2.33}Se_{0.67}$, $As_{0.33}SbBi_{0.67}S_{1.67}Te_{1.33}$ and $As_{0.33}Sb_{0.67}BiS_2Te$). In particular, the TIs ($Sb_{1.67}Bi_{0.33}S_{0.67}Te_{1.33}Se$ and $BiSbS_{0.67}SeTe_{1.33}$) as well as the NIs ($As_{0.33}SbBi_{0.67}S_{1.67}Te_{1.33}$ and $As_{0.33}Sb_{0.67}BiS_2Te$) are located in the region with $6806 < D_1 < 7854$ so that their topological character is determined by the descriptor $D_2$. For all the eight systems with fractional stoichiometry, the first-principles calculations of the $Z_2$ invariant fully confirm the predictions of the 2D descriptor. As an example, Fig. 4(a) shows the orbital-decomposed band structure of $Sb_2S_{0.33}Se_{0.67}Te_2$, in which a band inversion is present at the Γ point between the valence (mainly occupied by the $p_z$ orbitals of the Sb atoms) and conduction bands (mainly occupied by the $p_z$ orbitals of the Te atoms). Conversely, the band structure of SbBiS$_{2.33}$Se$_{0.67}$ shown in Fig. 4(b) exhibits a semiconducting behavior with normal band order. The crystal structures, band structures, and the corresponding evolution of Wannier centers for these eight tetradymites are summarized in Section 5 of the Supplemental Material [36]. Due to the moderate system sizes, these layered compounds necessarily exhibit pronounced short- (interlayer) and long- (intralayer) range orders, especially if compared with the much more disordered random alloys studied experimentally. Nonetheless, the ($D_1$, $D_2$) descriptor achieves excellent agreement between the SISSO predictions and first-principles calculations as well as the experimental data, further substantiating its reliable and robust nature in determining the topological characters of tetradymites with either integer or fractional stoichiometry.

## III. CONCLUSIONS

In summary, we have performed extensive, high-level first-principles calculations to obtain reliable information on the topological character of 243 tetradymites, of which many are discovered as new TIs. From this training data set, we have obtained a simple predictive 2D descriptor for the TI/NI classification using the AI-based SISSO approach. The descriptor only depends on the atomic number and Pauling electronegativity of the constituent elements, and captures the essential physics governing the TIs. Accordingly, it exhibits perfect classification accuracy and strong predictive power, even for systems drastically beyond the training data, as explicitly demonstrated for alloyed tetradymites compounds, i.e., a material class that can be created by alloying and would be prohibitively expensive to investigate even using the lowest-level first-principles approaches. With this descriptor, we are able to identify all the potential TIs in this complex material class in a fast and reliable fashion, most of which are out of the scope of the recently proposed symmetry-based indicators.

The approach established here should be applicable to the classification of the topological characters of many other classes of materials beyond the tetradymite family. The present study therefore offers a major step forward in exploration of the topological materials space, allowing us to investigate complex materials without prior determination of their specific symmetries and detailed band structures. Collectively, these findings also attest the increasingly important role of such AI-based approaches in modern materials discovery.

**ACKNOWLEDGMENTS**

The authors acknowledge fruitful discussions with Carlos Mera Acosta and Zizhen Zhou. We thank financial support from the National Natural Science Foundation of China (Grant Nos. 51772220, 11574236, 11634011, and 61434002), the National Key R&D Program of China (Grant No. 2017YFA0303500), and Anhui Initiative in Quantum Information Technologies (Grant No. AHY170000). Furthermore, this project was supported by TEC1p (the European Research Council (ERC) Horizon2020 research and innovation program, grant agreement No.740233), BigMax (the Max Planck Society's Research Network on Big-Data-Driven Materials-Science), and the NOMAD pillar of the FAIR-DI e.V. association. The numerical calculations in this work have been done on the platform in the Supercomputing Center of Wuhan University.

**FIGURES AND CAPTIONS**

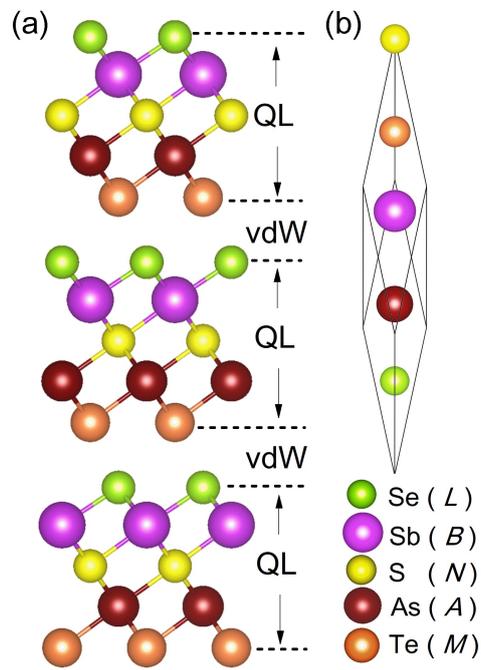

FIG. 1. Crystal structure of the quinary tetradymite AsSbSeTeS, where (a) is the unit cell, (b) is the primitive cell, and the atoms As, Sb, Se, Te, S occupy the sites $A$, $B$, $L$, $M$, $N$, respectively.

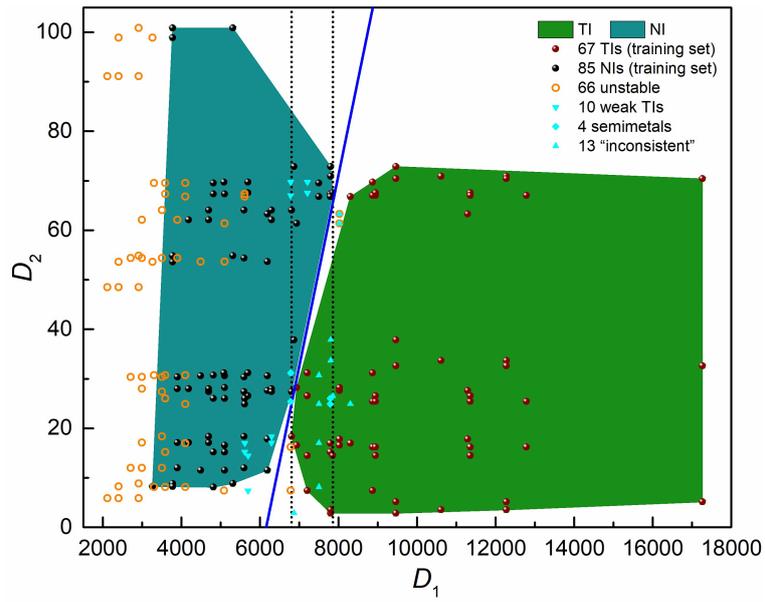

FIG. 2. Phase diagram for the 152 tetradymites used as the training data, plotted as a function of the 2D descriptor. The NI and TI phases are marked by the cyan and green areas, determined by connecting the outermost black and red points, respectively. For comparison, 91 additional tetradymites not included in the training set (see text) are also included in different colors.

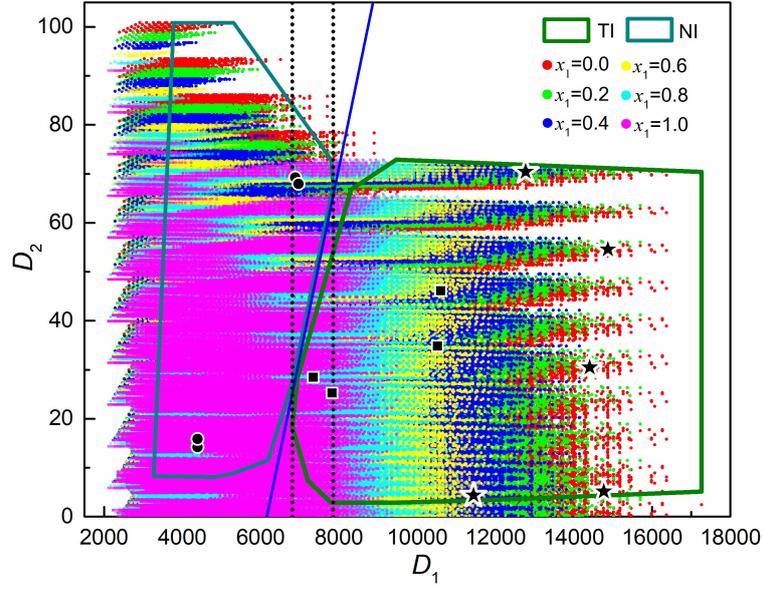

FIG. 3. A total of 4,084,101 possible tetradymites with fractional stoichiometries ($As_xSb_yBi_{2-x-y}S_aSe_bTe_{3-a-b}$) are mapped into the phase diagram using the calculated ($D_1$, $D_2$) values. The solid asterisks indicate 5 experimentally reported TIs with fractional stoichiometries. The topological characters of the 8 tetradymites marked by the solid circles (4 NIs) and solid squares (4 TIs) are confirmed by first-principles calculations.

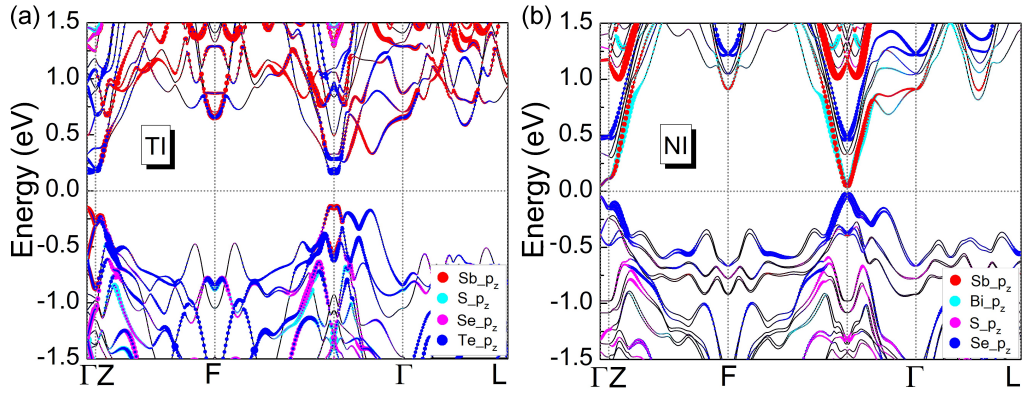

FIG. 4. Electronic band structures of (a) $Sb_2S_{0.33}Se_{0.67}Te_2$ and (b) $SbBiS_{2.33}Se_{0.67}$, which were calculated by using density functional theory in its projector augmented wave formulation. The Fermi level is at 0 eV. The colored circles denote the contributions from different orbitals, and the sizes of the markers are proportional to the magnitudes of their contributions.

*e-mails: phlhj@whu.edu.cn; carbogno@fhi-berlin.mpg.de; zhangzy@ustc.edu.cn